\documentclass[aps,pra,reprint,twocolumn,superscriptaddress,amsmath,amssymb]{revtex4-2}

\usepackage{verbatim}

\usepackage{array}
\usepackage{physics}
\usepackage{amsmath}  
\usepackage{amsfonts} 
\usepackage{graphicx} 
\usepackage{ragged2e}
\usepackage{color}
\usepackage{chemformula}
\usepackage[colorlinks=true,citecolor=blue,urlcolor=blue]{hyperref}
\usepackage{dcolumn}
\usepackage{bm}
\usepackage{orcidlink}

\begin{document}

\title{Observable Resonances in Efimov-unfavored Systems}

\author{Karim I. Elghazawy\,\orcidlink{0009-0003-4383-1939}}
\affiliation{Department of Physics and Astronomy, Purdue University, West Lafayette, Indiana 47907, USA}
\author{Chris H. Greene\,\orcidlink{0000-0002-2096-6385}}
\affiliation{Department of Physics and Astronomy, Purdue University, West Lafayette, Indiana 47907, USA}
\affiliation{\mbox{Purdue Quantum Science and Engineering Institute, Purdue University, West Lafayette, Indiana 47907, USA}}



\date{\today}

\begin{abstract} 
Three-body loss resonances associated with heavy-heavy-light Efimov states have been observed for over a decade in ultracold mixtures tuned near interspecies Feshbach resonances. For light-light-heavy systems, observing such resonances has been far more challenging due to the substantially large Efimov spacing. In these Efimov-unfavored systems, the intraspecies scattering length $a_\text{BB}$ has been shown to significantly affect the overall Efimov scenario, namely, the positions of the Efimov resonances $a_{-}^{(n)}$ and the three-body parameter (3BP) $a_{-}^{(0)}$. The present article explains the origin behind this influence by highlighting two primary mechanisms via which both the magnitude and sign of $a_\text{BB}$ govern the Efimov spectrum and set the resulting 3BP $a_{-}^{(0)}$. By employing van der Waals interactions for \ch{^{23}Na2^{40}K}, we attribute the vital role of $a_\text{BB}$ in Efimov-unfavored systems to the large difference between the Efimov scaling parameters for two and three resonant interactions, $s_0$ and $s_0^*$. In particular, we account for the unusually large $a_{-}^{(0)}$ obtained in light-light-heavy systems with $a_\text{BB}>0$ (e.g., \ch{^{41}K2^{87}Rb}), and show that the first Efimov resonance can still occur at an experimentally accessible value when $a_\text{BB}<0$.
\end{abstract}
    
\maketitle 

\section{Introduction} 
In the 1970s, Efimov predicted the existence of an infinite series of weakly bound three-body states (Efimov trimers) in systems with resonant two-body interactions, where the $s$-wave scattering length $a$ vastly exceeds the range of the pairwise interaction \cite{EFIMOV1970563}. A hallmark of this remarkably bizarre quantum effect is the constant ratio between successive bound-state energies: $E_n/E_{n+1}=e^{2 \pi / s_0}$, where $s_0$ is known as the Efimov scaling parameter, giving rise to a geometric series that accumulates at the three-body threshold. This behavior originates from a scale-invariant three-body attractive potential, $U(R) \propto -(s_0^2+1/4)/R^2$, that emerges in the unitary limit $|a|\rightarrow \infty$, where the hyperradius $R$ characterizes the average size of the system. The Efimov effect and its significance for ultracold few-body physics have been comprehensively studied and reviewed over the past few decades \cite{BRAATENReview,GreeneRevModPhys,NaidonReview2017}. 

The ability to fully tune the interactions strength---i.e., vary the scattering length $a$---using an external magnetic field \cite{RevModPhysFeshbach,6LiFeshbach} in systems supporting Feshbach resonances has enabled ultracold experiments to access the regime of the Efimov effect. Subsequently, theoretical interest in Efimov physics has grown substantially \cite{Petrov,Hammer2007,vonStecher2009,Wang2009,Zenesini_2013}, owing to its crucial role in understanding and controlling atomic losses in ultracold experiments. Efimov states have been extensively observed in homonuclear systems \cite{Kraemer2006Cs,Knoop2009,Berninger2011,HuangCs2014,Wild2012Rb,Zaccanti2009K,Pollack2009Li7,Gross2009_7Li,Ottenstein2008,Huckans2009_6Li,Wenz2009_6Li,Williams2009,Lompe2010,Nakajima2010,Ferlaino2009}, as $a$ is varied across a Feshbach resonance, through their signatures in the three-body loss rate $K_3$, appearing as resonances for $a<0$ and minima for $a>0$. 
Efimov resonances occur in $K_3$ at negative scattering lengths, denoted $a_{-}^{(n)}$, when distinct Efimov trimers reach the three-body threshold, forming a shape resonance that serves as an enhancing intermediate state for ultracold three-body recombination (3BR) \cite{Esry99}. 
    
Although the Efimov series extends infinitely toward the three-body threshold, it is bounded from below by the physics at short interparticle distances through a three-body parameter (3BP) \cite{BRAATENReview}. This parameter encodes crucial information about the system's ultracold behavior, as it governs the weakly bound Efimov spectrum and determines scattering properties at low collision energy such as cross sections and event rates. Intuitively, the 3BP can be viewed as the initial element of the Efimov geometric series, which, together with the ratio between successive elements, fully specifies the series. Common definitions of the 3BP include the ground state energy at unitarity, $E_0$, and the scattering length associated with the first Efimov resonance, $a_-^{(0)}$. Throughout this article, the latter definition of the 3BP is adopted.

\begin{figure}[b!]
    \centering
    \includegraphics[width=0.9\linewidth]{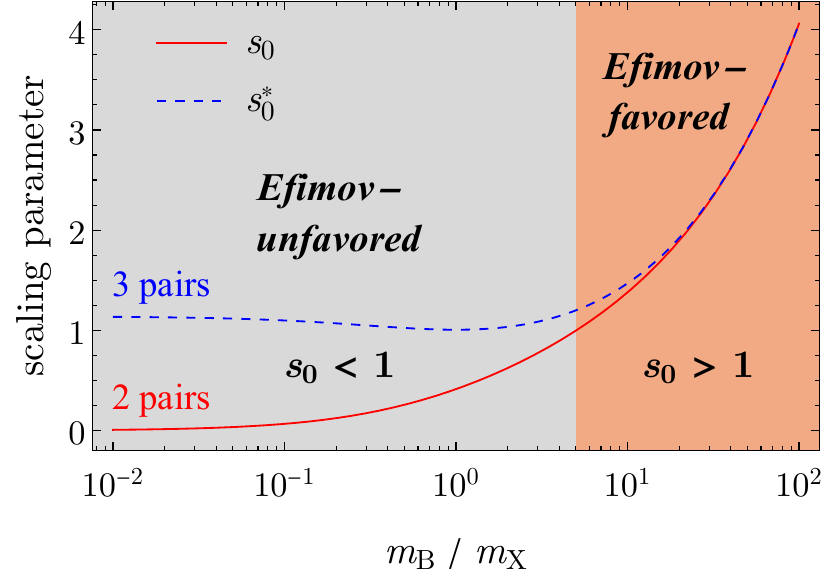}
    \caption{The Efimov scaling parameters for two (solid) and three (dashed) resonant pairs $s_0$ and $s_0^*$ plotted vs. the system mass imbalance $m_\text{B}/m_\text{X}$}
    \label{fig0sp}
\end{figure}
While the Efimov effect is primarily known for three identical atoms, it also occurs in heteronuclear systems (BBX) composed of two identical bosons and a distinguishable atom, provided that at least two of the pairwise interactions are resonant (i.e., the interspecies scattering length $\abs{a_\text{BX}}\rightarrow\infty$) \cite{EFIMOV1973157,D'Incao2006,Helfrich2010,GreeneHetero,GreeneHeteroErratum,Petrov2015}. In these systems, two Efimov scaling parameters arise depending on whether only two ($s_0$) or all three ($s_0^*$) interactions are resonant \cite{NaidonReview2017}, both universally determined by the system's mass imbalance (see Fig.~\ref{fig0sp}). Efimov resonances $a_{-}^{(n)}$ have been reported in ultracold two-species gases \cite{Barontini2009KRb,Wacker2016RbK,Maier2015RbLi,Tung2014CsLi,Ulmanis2016CsLi1} as $a_\text{BX}$ is varied near an interspecies Feshbach resonance, thereby measuring the 3BP $a_-^{(0)}$. As depicted in Fig.~\ref{fig0sp}, ``Efimov-favored" systems, involving two very heavy and one light atom ($m_\text{B}/m_\text{X}\gtrsim5$), feature a relatively small Efimov spacing  ($s_0>1$), enabling the detection of multiple states in the geometric series \cite{Tung2014CsLi,Ulmanis2016CsLi1}, in contrast to ``Efimov-unfavored" systems ($s_0<1$). Heteronuclear trimers have also been more directly observed via radio-frequency association, which simultaneously provided measurements of their binding energies \cite{Zwierlein}.

In early studies, the intraspecies scattering length $a_\text{BB}$ was treated as a background parameter, with no significant influence on the Efimov resonances $a_{-}^{(n)}$ or, in particular, on the 3BP. For instance, Wang \textit{et al.} \cite{GreeneHetero,GreeneHeteroErratum} calculated the 3BP for various mass ratios, each at a single positive $a_\text{BB}$, finding excessively large values of $|a_{-}^{(0)}|$ (e.g., $10^6\,a_0$) in the Efimov-unfavored cases. Later, evidence of this influence was revealed in the \ch{Cs2Li} system, where atomic losses were examined near two interspecies Feshbach resonances with opposite signs of $a_\text{CsCs}$ \cite{Ulmanis2016CsLi2,Hafner2017CsLi}. In the case of $a_\text{CsCs}>0$, the lowest expected Efimov resonance was not present, as the \ch{Cs2} + \ch{Li} threshold prevents the corresponding Efimov state from reaching the three-body continuum, resulting in $a_{-}^{(0)}$ being nearly an order of magnitude larger than for $a_\text{CsCs}<0$. In Efimov-unfavored systems, however, $a_\text{BB}$ is expected to play a more prominent role, as the large discrepancy between $s_0$ and $s_0^*$, displayed in Fig.~\ref{fig0sp}, underscores the significance of the intraspecies interaction, which controls the shift between the regimes of two and three resonant pairs. For example, the effect of $|a_\text{BB}|$ on recombination peaks \cite{Mikkelsen_2015} and on the 3BP $a_-^{(0)}$ \cite{Zhao_HHL} has been theoretically examined for $a_\text{BB}<0$ in different heavy-heavy-light mixtures, where markedly stronger variations were found in systems with smaller mass imbalance, such as \ch{Rb2K} and \ch{Rb2Na}, which are Efimov-unfavored. 

This work elucidates the physical origin underlying the strong effect of $a_\text{BB}$ on both the 3BP $a_{-}^{(0)}$ and the resonance positions $a_{-}^{(n)}$ in Efimov-unfavored systems. Specifically, we identify distinct mechanisms through which the Efimov spectrum (and the associated $a_{-}^{(n)}$) is controlled by $a_\text{BB}$, depending on its sign. Pairwise van der Waals interactions modeled with separable potentials are utilized for \ch{^{23}Na2^{40}K}, a newly observed example \cite{Zwierlein}, to calculate the unitarity energy spectrum ($\abs{a_\text{BX}}\rightarrow\infty$) and the resonance positions $a_{-}^{(n)}$ across the universal range in $a_\text{BB}$. For $a_\text{BB}<0$, a strong variation of the unitarity spectrum manifests in the intermediate regime, where the homonuclear pair becomes ``partially" resonant as the system transitions between three and two resonant pairs. Conversely, for $a_\text{BB}>0$, the homonuclear dimer threshold splits the three-body spectrum into two widely spaced (in energy) Efimov spectra. Accordingly, a similar, yet dramatically more pronounced effect than in the \ch{Cs2Li} system \cite{Ulmanis2016CsLi2,Hafner2017CsLi} leads to the absence of an entire series of Efimov resonances, rather than just one, yielding a significantly larger 3BP $a_{-}^{(0)}$ (by five orders of magnitude) compared to the case of $a_\text{BB}<0$. This provides an explanation for the large values of $|a_{-}^{(0)}|$ found for Efimov-unfavored mixtures with $a_\text{BB}>0$ \cite{GreeneHetero,GreeneHeteroErratum} and demonstrates that the 3BP can attain a measurable value in light-light-heavy (LLH) systems, provided $a_\text{BB}<0$. These mechanisms, revealed by the finite-range model, are interpreted and supported by zero-range hyperspherical potential curves. Additionally, we theoretically reproduce the measured binding energy of the \ch{^{23}Na2^{40}K} trimer recently reported \cite{Zwierlein} and highlight its position within the overall Efimov spectrum.

\section{Theory}
\subsection{Three-Body Bound States}  
The Schr\"odinger equation in momentum space for three distinguishable particles reads, after removal of the center of mass motion,
\begin{multline}
    \left(\frac{\hbar^2 p_i^2}{2 \mu^i}+ \frac{\hbar^2 q_i^2}{2 \mu_i}\right)\Psi(\vec{p}_i,\vec{q}_i) \ + \\ \sum_{j=1}^{3}\int\frac{d^3\vec{q}_j\!'}{(2\pi)^3} V_j(\vec{q}_j,\vec{q}_j\!') \Psi(\vec{p}_j,\vec{q}_j\!')=
     E\, \Psi(\vec{p}_i,\vec{q}_i), \label{threebody}
\end{multline}
where the Jacobi coordinate set $(\vec{p}_i,\vec{q}_i)$ is employed to express the kinetic term, while the appropriate set $(\vec{p}_j,\vec{q}_j)$ is used for each pairwise interaction term. The Jacobi momenta and corresponding reduced masses $(\mu^i,\mu_i)$ are defined in Appendix~\ref{appendixA}. 
The two-body interactions are assumed to be described by separable potentials \cite{Yamaguchi_Separable}: 
\begin{equation}
    V_i(\vec{q},\vec{q}\,') = \frac{\hbar^2}{2 \mu_i}\alpha_i \ \chi_i(\vec{q}\,) \chi^*_i(\vec{q}\,'). \label{twobody}
\end{equation}
Inserting these separable potentials into Eq.~(\ref{threebody}) gives
\begin{equation}
        \left(\frac{p_i^2}{\mu^i}+ \frac{q_i^2}{\mu_i} - \frac{2E}{\hbar^2}\right) \Psi(\vec{p}_i,\vec{q}_i) + \sum_{j=1}^{3} \chi_j(\vec{q}_j) F_j(\vec{p}_j)= 0 \label{three},
\end{equation}
with the unknown functions $F_i(\vec{p}_i)$ defined as
\begin{equation}
     F_i(\vec{p}_i) \equiv \frac{\alpha_i}{\mu_i} \int\frac{d^3\vec{q}_i}{(2\pi)^3} \chi^*_i(\vec{q}_i) \Psi(\vec{p}_i,\vec{q}_i). \label{F_def}
\end{equation}
Eq.~(\ref{three}) can be solved for $\Psi(\vec{p}_i,\vec{q}_i)$ as 
\begin{equation}
   \Psi(\vec{p}_i,\vec{q}_i) = - \sum_{j=1}^{3} \frac{\chi_j(\vec{q}_j) F_j(\vec{p}_j)}{\frac{p_i^2}{\mu^i}+ \frac{q_i^2}{\mu_i} - \frac{2E}{\hbar^2}} \label{psisoln}.
\end{equation}
Note that, since bound-state solutions ($E<0$) are of interest here, the right-hand side of Eq.~(\ref{psisoln}) includes only a particular solution term and no homogeneous solution. By inserting Eq.~(\ref{psisoln}) into Eq.~(\ref{F_def}), the following system of coupled integral equations is obtained
\begin{multline}
    \Bigg(\frac{\mu_i}{\alpha_i}+\int\frac{d^3\vec{q}_i}{(2\pi)^3} \frac{\abs{\chi_i(\vec{q}_i)}^2}{\frac{p_i^2}{\mu^i}+ \frac{q_i^2}{\mu_i} - \frac{2E}{\hbar^2}}\Bigg) F_i(\vec{p}_i) \; + \\ \sum_{j\neq i}\int\frac{d^3\vec{q}_i}{(2\pi)^3} \frac{\chi^*_i(\vec{q}_i) \chi_j(\vec{q}_j)}{\frac{p_i^2}{\mu^i}+ \frac{q_i^2}{\mu_i} - \frac{2E}{\hbar^2}}F_j(\vec{p}_j)=0. \label{IntEqs1}
\end{multline}
Now, this system is not readily solvable numerically since the argument of $F_j(\vec{p}_j)$ differs from the integration variable $\vec{q}_i$, which is essential for recasting the integral operator as a matrix multiplication. To this purpose, the following relations (derivable from Eqs.~(\ref{Jacobi_def})) are used to write the Jacobi set $(\vec{p}_j,\vec{q}_j)$ in terms of $(\vec{p}_i,\vec{q}_i)$
\begin{equation}
\begin{aligned}
     \vec{p}_j &= \frac{-m_j}{m_j+m_k} \, \vec{p}_i \; + \; \epsilon_{ijk} \, \vec{q}_i \\
     \vec{q}_j &= \epsilon_{jik} \frac{m_k M}{(m_i+m_k)(m_j+m_k)} \, \vec{p}_i \; + \; \frac{-m_i}{m_i+m_k} \, \vec{q}_i,
\end{aligned}
\end{equation}
where $(i,j,k)$ form a permutation of $(1,2,3)$, $\epsilon_{ijk}$ is the Levi-Civita symbol, and $M$ is the total mass. Upon plugging these expressions into Eq.~(\ref{IntEqs1}) and redefining the integration variable as $\vec{q}_i \rightarrow \epsilon_{ijk}\left(\vec{q}_i+\frac{m_j}{m_j+m_k} \, \vec{p}_i\right)$, one gets
\begin{multline}
    \Bigg(\frac{\mu_i}{\alpha_i}+\int\frac{d^3\vec{q}}{(2\pi)^3} \frac{\abs{\chi_i(\vec{q}\,)}^2}{\frac{p^2}{\mu^i}+ \frac{q^2}{\mu_i} - \frac{2E}{\hbar^2}}\Bigg) F_i(\vec{p}\,) \; + \sum_{j\neq i} \int\frac{d^3\vec{q}}{(2\pi)^3} \\ \times \frac{\chi^*_i\!\left(\epsilon_{ijk}\left[\vec{q}+\frac{m_j}{m_j+m_k} \vec{p}\,\right]\right) \chi_j\!\left(\epsilon_{jik}\left[\vec{p}+\frac{m_i}{m_i+m_k}\vec{q}\,\right]\right)}{\frac{p^2}{\mu_j}+ \frac{q^2}{\mu_i}+\frac{2}{m_k}\vec{p}\cdot\vec{q} - \frac{2E}{\hbar^2}} \\ \times F_j(\vec{q}\,)=0, \label{IntEqs2}
\end{multline}
where the indices on the momentum coordinates $(\vec{p},\vec{q}\,)$ have been omitted.

Thus far, no presumptions have been made on the interaction form factors $\chi_i(\vec{q}\,)$. Moving forward, $s$-wave interactions are assumed among the three particles, i.e., $\chi_i(\vec{q}\,)=\chi_i(q)$. Following Naidon, Endo, and Ueda \cite{Naidon2014}, our treatment adopts the following form factor that models van der Waals pairwise interactions
\begin{equation}
    \chi_i(q) = 1 - q\int_0^\infty\bigg(1-\frac{r}{a_i}-\varphi_i(r)\bigg)\sin(q r) \,dr,
\end{equation}
and the prefactor $\alpha_i$ is selected as
\begin{equation}
      \frac{1}{\alpha_i} = \frac{1}{4\pi a_i} - \frac{1}{2\pi^2} \int_0^\infty dq\abs{\chi_i(q)}^2, \label{alpha}
\end{equation}
where $a_i$ are the $s$-wave scattering lengths and $\varphi(r)$ denotes the zero-energy $s$-wave radial solution of the van der Waals potential in the two-body sector. This form factor choice is shown to reproduce the low-energy properties of the full potential, including the scattering length and the shallow dimer bound state. In particular, it yields the exact solution of the two-body Schr\"odinger equation at zero energy, namely $\varphi(r)$.  Consequently, it is important to note that this form factor is numerically reliable for scattering lengths $|a_i|\gg\ell_i^\text{vdW}$, where $\ell_i^\text{vdW}=\frac12(2\mu_i C_{6,i}/ \hbar^2)^{1/4}$ is the van der Waals length of the corresponding interaction, and its accuracy diminishes progressively as one moves outside this regime.

For three-body states with total angular momentum $L$, the angular dependence of $F_i(\vec{p}\,)$ is given by $F_i(p) \, Y_{LM}(\hat{p})$. Utilizing the previous relations, Eq.~(\ref{IntEqs2}) is reduced, after multiplying by $Y^*_{LM}(\hat{p})$ and summing over $M$, to a one-dimensional system of coupled integral equations:
\begin{equation}
    D_i(p)F_i(p)+\sum_{j\neq i}\int_0^\infty H_{ij}(p,q)F_j(q)\,dq=0, \label{IntEqs}
\end{equation}
with
\begin{multline}
    D_i(p) = \mu_i \Bigg(\frac{\pi}{a_i}-2\int_0^\infty dq\frac{\big(\frac{p^2}{\mu^i}-\frac{2E}{\hbar^2}\big)\abs{\chi_i(q)}^2}{\frac{q^2}{\mu_i}+\frac{p^2}{\mu^i} - \frac{2E}{\hbar^2}}\Bigg),
\end{multline}
\begin{multline}
    H_{ij}(p,q) = \int_{-1}^1 du \, P_L(u) \, q^2 \\ \times \frac{\chi^*_i\!\left(\abs{\vec{q}+\frac{m_j}{m_j+m_k} \vec{p}\,}\right) \chi_j\!\left(\abs{\vec{p}+\frac{m_i}{m_i+m_k}\vec{q}\,}\right)}{\frac{p^2}{\mu_j}+ \frac{q^2}{\mu_i}+\frac{2}{m_k}p q u- \frac{2E}{\hbar^2}} ,
\end{multline}
where $P_L(u)$ is a Legendre polynomial, and $\vec{p}\cdot \vec{q} = pqu$. By discretizing the linear operators in Eqs.~(\ref{IntEqs}), one can solve for the bound-state energy $E$ by searching for the roots of the resulting determinantal equation. Stated differently, Eqs.~(\ref{IntEqs}) can be recast as a $3\times3$ block-matrix operator acting on the vector of unknown functions $F_i$ as follows:
\begin{equation}
    \begin{pmatrix}
    D_1 & H_{12} & H_{13}\\
    H_{21} & D_2 & H_{23}\\
    H_{31} & H_{32} & D_3
    \end{pmatrix}
    \begin{pmatrix}
    F_1 \\ F_2 \\ F_3
    \end{pmatrix} 
    =0, \label{DisIntEq}
\end{equation}
where each element of the block matrix is itself a matrix, tabulated on a grid of momentum values $(p,q)$. Hence, by seeking the special energy values at which one of the eigenvalues (and the determinant) of this block matrix vanishes, the allowed trimer energies are determined for any masses $m_i$, $s$-wave scattering lengths $a_i$, and total angular momentum $L$.

Note that instead of searching for roots in energy $E$ at a fixed set of scattering lengths $a_i$, one can search for roots in scattering lengths at a given energy. For example, this approach proves useful for determining the resonance positions $a_-^{(n)}$ in both homonuclear and heteronuclear systems by solving the determinantal equation at the three-body threshold ($E=0$).

\subsection{Imposing Particles Symmetry}
In this subsection, the symmetry constraints are derived for systems with arbitrary exchange symmetry (i.e., some or all of the three particles are identical bosons/fermions). These constraints, which will ultimately impose conditions on the unknowns $F_i(p)$, simplify the system in Eqs.~(\ref{IntEqs}) by reducing the total number of equations and unknowns. 

The starting point is the fact that permuting the relevant bosons (fermions), along with their intrinsic degrees of freedom, should not affect (or should introduce a minus sign to) the properly symmetrized (antisymmetrized) three-body wavefunction. For example, if particles 1 and 2 are identical bosons, permuting their momenta ($\vec{k}_1$ and $\vec{k}_2$) should leave the wavefunction unchanged. Given that $\hat{P}_{12} (\vec{p}_1,\vec{q}_1) = (\vec{p}_2,-\vec{q}_2)$ and $\hat{P}_{12} (\vec{p}_3,\vec{q}_3) = (\vec{p}_3,-\vec{q}_3)$, Eq.~(\ref{psisoln}) implies that
\begin{align} 
\label{symm12}
    \nonumber    &\chi_1(\vec{q}_1) F_1(\vec{p}_1) + \chi_2(\vec{q}_2) F_2(\vec{p}_2) + \chi_3(\vec{q}_3) F_3(\vec{p}_3) = \\ 
    \nonumber    &\chi_1(-\vec{q}_2) F_1(\vec{p}_2) + \chi_2(-\vec{q}_1) F_2(\vec{p}_1) + \chi_3(-\vec{q}_3) F_3(\vec{p}_3) = \\
    &\chi_2(\vec{q}_2) F_1(\vec{p}_2) + \chi_1(\vec{q}_1) F_2(\vec{p}_1) + \chi_3(\vec{q}_3) F_3(\vec{p}_3),
\end{align}
where the property of $s$-wave interactions $\chi_i(-\vec{q}\,)=\chi_i(\vec{q}\,)$ and the fact that $\chi_1 = \chi_2$, following from particles 1 and 2 being identical, were used in the last equality. Comparing the first and last lines of Eq.~(\ref{symm12}), leads to the conclusion that $F_2=F_1$ and no restriction on $F_3$, simplifying the original coupled system in Eqs.~(\ref{IntEqs}) to two equations in two unknowns $F_1(\vec{p})$ and $F_3(\vec{p})$. Similarly, if particles 1 and 2 are fermions, one could show that the required symmetry constraints are $F_2 = - F_1$ and $F_3=0$. The vanishing of $F_3$ in this case merely reflects the fact that fermions cannot interact via $s$-wave interactions. However, if a $p$-wave interaction is allowed between only the two fermions, i.e., $ \chi_3(-\vec{q}_3)= -\chi_3(\vec{q}_3)$, one finds that $F_3 \neq 0$. The required exchange-symmetry constraints on $F_i$ for different systems are summarized in Table~\ref{symmtable}.

\begin{table}[ht!]
\centering
\caption{Symmetry constraints are provided for three-body systems with different exchange symmetries (assuming $s$-wave interactions), where X denotes a distinguishable particle, B an identical boson, and F an identical fermion.} \vspace{2pt} 
\begin{tabular}{c r r c}
   \hline\hline
    \hspace{19pt} XXX \hspace{19pt} & \hspace{18pt} $F_1$ \hspace{18pt} & \hspace{18pt} $F_2$ \hspace{18pt} & \hspace{18pt} $F_3$ \hspace{18pt} \\ 
    \hline
    \hspace{19pt} BBX \hspace{19pt} & \hspace{18pt} $F_1$ \hspace{18pt} & \hspace{18pt} $F_1$ \hspace{18pt} & \hspace{18pt} $F_3$ \hspace{18pt} \\
    \hspace{19pt} FFX \hspace{19pt} & \hspace{18pt} $F_1$ \hspace{18pt} & \hspace{18pt} $-F_1$ \hspace{18pt} & \hspace{18pt} $0$ \hspace{18pt} \\
    \hspace{19pt} BBB \hspace{19pt} & \hspace{18pt} $F_1$ \hspace{18pt} & \hspace{18pt} $F_1$ \hspace{18pt} & \hspace{18pt} $F_1$ \hspace{18pt} \\
    \hline\hline
\end{tabular}
\label{symmtable}
\end{table} 

\section{Results and Discussion} \label{results}
In this section, the outcomes of computations carried out for \ch{^{23}Na2^{40}K}, an Efimov-unfavored system, are presented and discussed. The two van der Waals lengths used for this system are $\ell_\text{BB}=45 \,a_0$ and $\ell_\text{BX}=53.6\,a_0$ \cite{C6_2014,C6_2003}. This system has a BBX exchange symmetry with $a_1=a_2\equiv a_\text{BX}$ and $a_3 \equiv a_\text{BB}$. Applying the appropriate constraint (see Table~\ref{symmtable}) reduces Eq.~(\ref{DisIntEq}) to a $2\times2$ system, which is then solved for spherically symmetric trimer states ($L=0$) over a range of values of one scattering length ($a_{\text{BB}}$ or $a_{\text{BX}}$) with the other held fixed. As mentioned earlier, the same system can also be directly solved for the values $a_\text{BX}=a_-^{(n)}$ by setting the total energy to $E = 0$ and keeping $a_\text{BB}$ fixed. Recall that the separable potential model employed here reproduces exactly the two-body interactions only at zero energy. Therefore, quantitative accuracy can be expected primarily for $|a_\text{BX}|\gg\ell_\text{BX}$ and $|a_\text{BB}|\gg\ell_\text{BB}$. Results associated with Eqs.~(\ref{IntEqs}) are henceforth referred to as finite-range (FR) calculations. 

The results of FR calculations in momentum space are compared with those from real-space calculations based on three-body adiabatic potential curves with contact $s$-wave interactions. The zero-range hyperspherical potentials for three particles are given by \cite{GreeneZR}
\begin{equation}
    U_n(R)=\frac{\hbar^2}{2\mu}\left[\frac{\lambda_n(R)-1/4}{R^2}-Q_{nn}(R)\right], \label{ZRpot}
\end{equation}
where $Q_{nn}$ represents the nonadiabatic diagonal corrections. The hyperangular eigenvalues $\lambda_n$ are computed as the determinantal roots (at each $R$) of a $3\times3$ matrix $\mathbf{Z}^{L,R}_{m_i,a_i}(\lambda)$ ($2\times2$ for BBX), whose elements are known analytically for any masses $m_i$, scattering lengths $a_i$, and total angular momentum $L$ (see Appendix~\ref{appendixB}). Upon generating the pertinent potential curves, one can look for three-body bound states associated with each adiabatic potential. Results derived from this model are subsequently referred to as zero-range (ZR) calculations.

\begin{figure}[b]
    \centering
    \includegraphics[width=\linewidth]{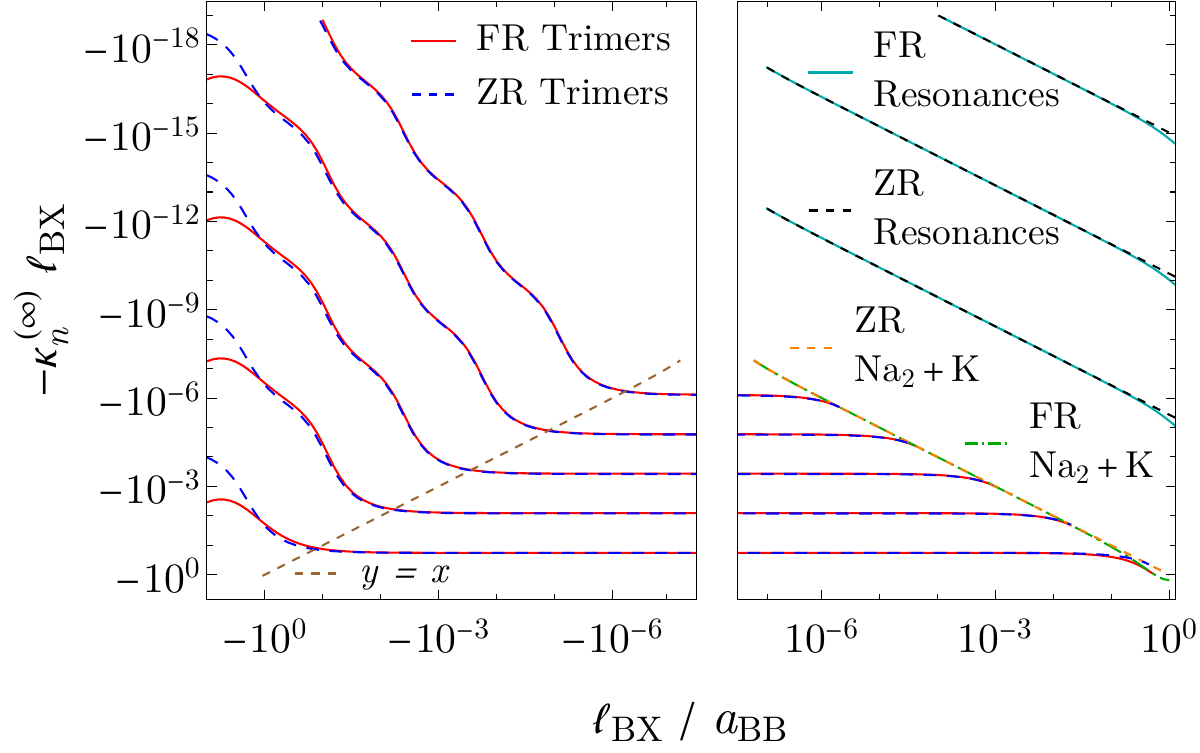}
    \caption{The Efimov spectrum for \ch{^{23}Na2^{40}K} at heteronuclear unitarity ($\abs{a_\text{BX}}\rightarrow\infty$) with $\kappa_n=\sqrt{-m_\text{B}E_n}/\hbar$. Each plotted quantity is computed from both a finite-range and a zero-range calculation. The dimer threshold (green and orange) divides the spectrum into lower trimers (red and blue) and upper resonances (cyan and black).}
    \label{fig1spectrum}
\end{figure}
\begin{figure*}[t!]
    \centering
    \includegraphics[height=0.32\linewidth]{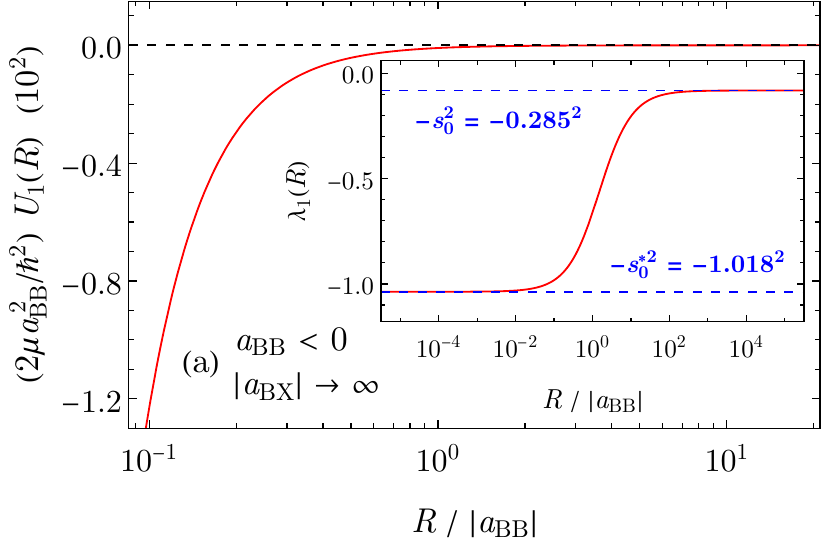}
    \hspace{14pt}
    \includegraphics[height=0.32\linewidth]{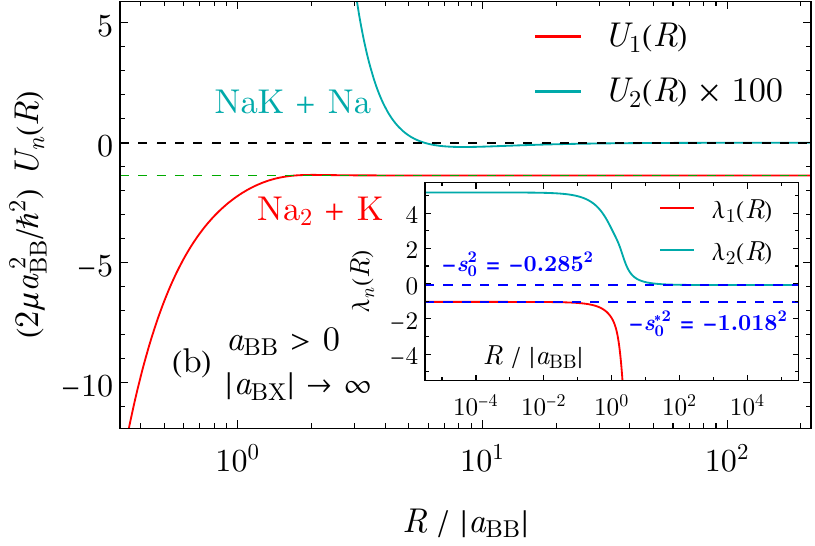}
    \caption{The zero-range hyperspherical potentials $U_n(R)$ at heteronuclear unitarity ($\abs{a_\text{BX}}\rightarrow\infty$), shown for $a_\text{BB}<0$ (a) and $a_\text{BB}>0$ (b). The insets display the corresponding hyperangular eigenvalues $\lambda_n(R)$ defined in Eq.~(\ref{ZRpot}). In the right panel, the upper potential $U_2(R)$ (cyan) is magnified by a factor of 100 for better visibility.}  
    \label{fig2potcurves}
\end{figure*}
\subsection{The Unitarity Spectrum}
The (rescaled) energies of the lowest several Efimov states at  $\abs{a_\text{BX}}\rightarrow\infty$, as well as the \ch{Na2} dimer threshold, are displayed in Fig.~\ref{fig1spectrum} as functions of $1/a_\text{BB}$. While solving the hyperradial Schr\"odinger equation with the ZR potential (Eq.~(\ref{ZRpot})) at $\abs{a_\text{BB}}\rightarrow\infty$, a short-range log-derivative was chosen to match an eigenvalue to the FR model's ground state energy. This log-derivative was then fixed for the ZR calculations at other $a_\text{BB}$ values. For $a_\text{BB}<0$, the spectrum follows Efimov geometric scaling in two main regions, with the ratios of successive energies being relatively small near $a_\text{BB}\rightarrow-\infty$ and gradually increasing to a much larger value as $a_\text{BB}\rightarrow0^-$. Moving leftward from $1/a_\text{BB}=0$, the Efimov states enter an intermediate region where their energies exhibit oscillatory behavior while progressively rising until the new scaling is established. The boundaries of this intermediate region differ for each state, as more excited states rise earlier than deeper ones.

The rationale behind the structure of the spectrum can be understood by examining the relevant ZR potential curves and their dependence on $a_\text{BB}$. As evident in the inset of Fig.~\ref{fig2potcurves}(a), for $a_\text{BB}<0$, the hyperangular eigenvalue $\lambda_1(R)$, which controls the Efimov scaling (see Eq.~(\ref{ZRpot})), takes on two constant values: $-s_0^{*2}$ as $R \ll \abs{a_\text{BB}}$ and $-s_0^2$ as $R \gg \abs{a_\text{BB}}$, with a connecting ``transition region", occurring at $R_0 \sim \abs{a_\text{BB}}$. This indicates the presence of a pure Efimov three-body potential as $a_\text{BB}\rightarrow-\infty$ (three resonant pairs) and as $a_\text{BB}\rightarrow0^-$ (two resonant pairs), featuring the series of Efimov states with the smaller ($s_0^*=1.018$) and larger spacings ($s_0=0.285$), respectively. 
In general, a bound state within the unitarity potential for $a_\text{BB}<0$ (Fig.~\ref{fig2potcurves}(a)) experiences a variable scaling parameter ranging from $s_0^*$ to $s_0$, depending significantly on the relative positions of the transition region $R_0$ and the antinode associated with the state's turning point $R_t \sim 1/\kappa_n^{(\infty)}$. For instance, when $a_\text{BB} \rightarrow -\infty$, the transition region is located at $R_0\rightarrow\infty$ ($R_0\gg R_t$), and all states see the $s_0^*$ scaling. As $\abs{a_\text{BB}}$ decreases, one Efimov state acquires a new scaling when the position of the transition region becomes comparable to its turning point ($R_0 \sim R_t$), causing the state to become less bound (i.e., rise) since it encounters an increase in potential energy. This occurs at $1/\abs{a_\text{BB}} \sim \kappa_n^{(\infty)}$, i.e., near the $y=x$ line depicted in Fig.~\ref{fig1spectrum} for $a_\text{BB}<0$. 
\begin{figure}[b!]
    \centering
    \includegraphics[width=\linewidth]{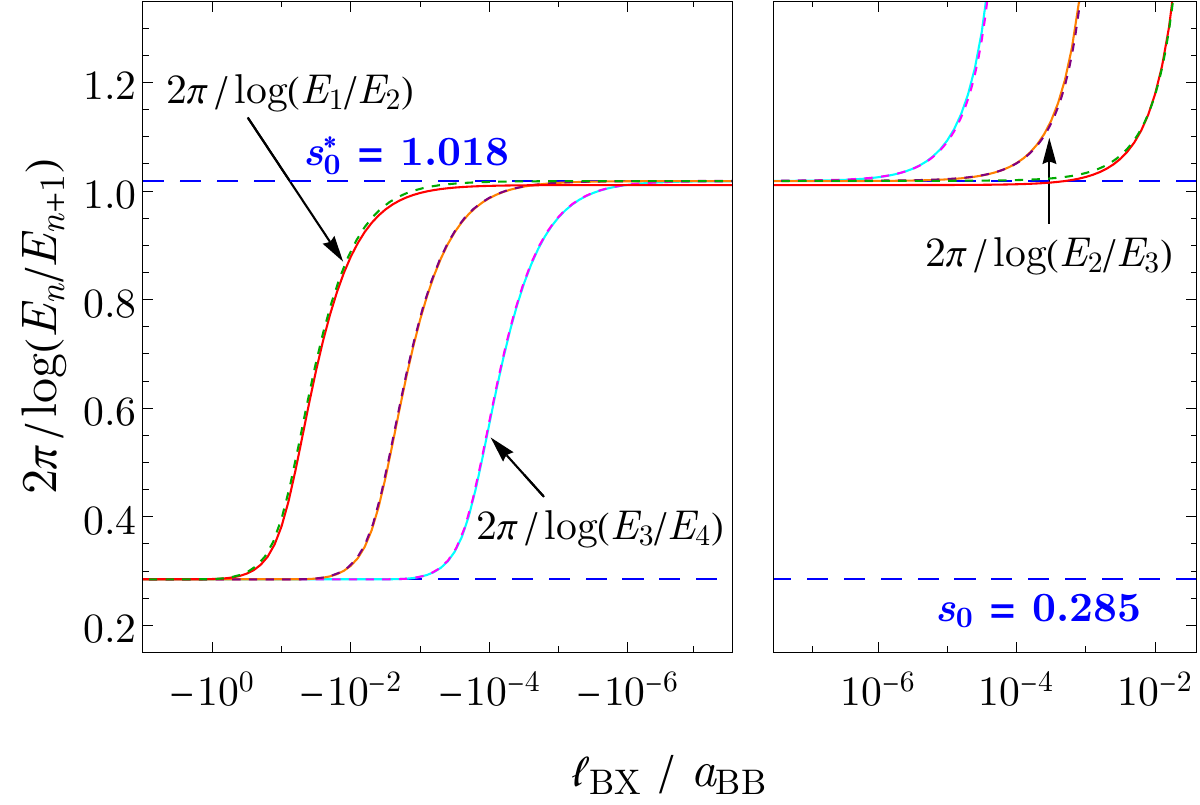}
    \caption{The variable Efimov scaling parameter, given through the ratios $E_n/E_{n+1}$ at $\abs{a_\text{BX}} \rightarrow\infty$, graphed for the lowest three pairs of consecutive trimer states in Fig.~\ref{fig1spectrum}. Solid curves correspond to FR calculations, while dashed curves correspond to ZR calculations. The horizontal dashed lines represent the universal Efimov scaling parameters: $s_0^*=1.018$ (upper) at $1/a_\text{BB}=0$ and $s_0=0.285$ (lower) at $1/a_\text{BB}\rightarrow-\infty$, corresponding to three and two resonant pairs respectively.}
    \label{fig3ratio}
\end{figure}
Moreover, the energy curve of each Efimov state displays one additional oscillation compared to the next lower state, since the wavefunctions of successive states contain an increasing number of antinodes, each overlapping with the transition region in a manner similar to the antinode at the turning point.
The variation of the scaling parameter, produced by the interplay between the transition region and the bound-state antinodes as $a_\text{BB}$ changes from $-\infty$ to $0$, is summarized in Fig.~\ref{fig3ratio}, which involves ratios of successive energies calculated from the unitarity spectrum. We emphasize, however, that the part of the unitarity spectrum (Fig.~\ref{fig1spectrum}) where $\ell_\text{BX}/a_\text{BB} \lesssim -1$ should be interpreted only qualitatively, as it lies outside the strict validity range of the FR model. Importantly,  this region is shown mainly to illustrate the transition between the $s_0$ and $s_0^*$ scalings.

In contrast to the Efimov spectrum for $a_\text{BB}<0$, the spectrum for $a_\text{BB}>0$ does not exhibit a smooth transition between two scaling laws. Instead, it is abruptly divided by the \ch{Na2} + \ch{K} threshold into two well-separated Efimov spectra having different scaling parameters: a lower spectrum (trimers) governed by $s_0^*$ and an upper spectrum (resonances) governed by $s_0$. This is supported by the shape of the adiabatic potentials for $a_\text{BB}>0$ presented in Fig.~\ref{fig2potcurves}(b). The lower and upper Efimov spectra reside in two extremely weakly coupled potentials, $U_1$ (red) and $U_2$ (cyan), which are tied to the \ch{Na2} + \ch{K} and \ch{NaK} + \ch{Na} channels, respectively. The inset of Fig.~\ref{fig2potcurves}(b) shows the corresponding hyperangular eigenvalues approaching different limits: $\lambda_1 \rightarrow-s_0^{*2}$ as $R \ll \abs{a_\text{BB}}$, while $\lambda_2\rightarrow -s_0^{2}$ as $R \gg \abs{a_\text{BB}}$, indicating that each Efimov spectrum is controlled by a distinct scaling parameter. This effect becomes more pronounced as $m_\text{B}/m_\text{X}$ decreases, since the separation between the potentials $U_1$ and $U_2$, and consequently the corresponding spectra, is governed by the difference between $s_0$ and $s_0^*$ (see Fig.~\ref{fig0sp} and the inset of Fig.~\ref{fig2potcurves}(b)). 
\begin{figure}[b!]
    \centering
\includegraphics[width=\linewidth]{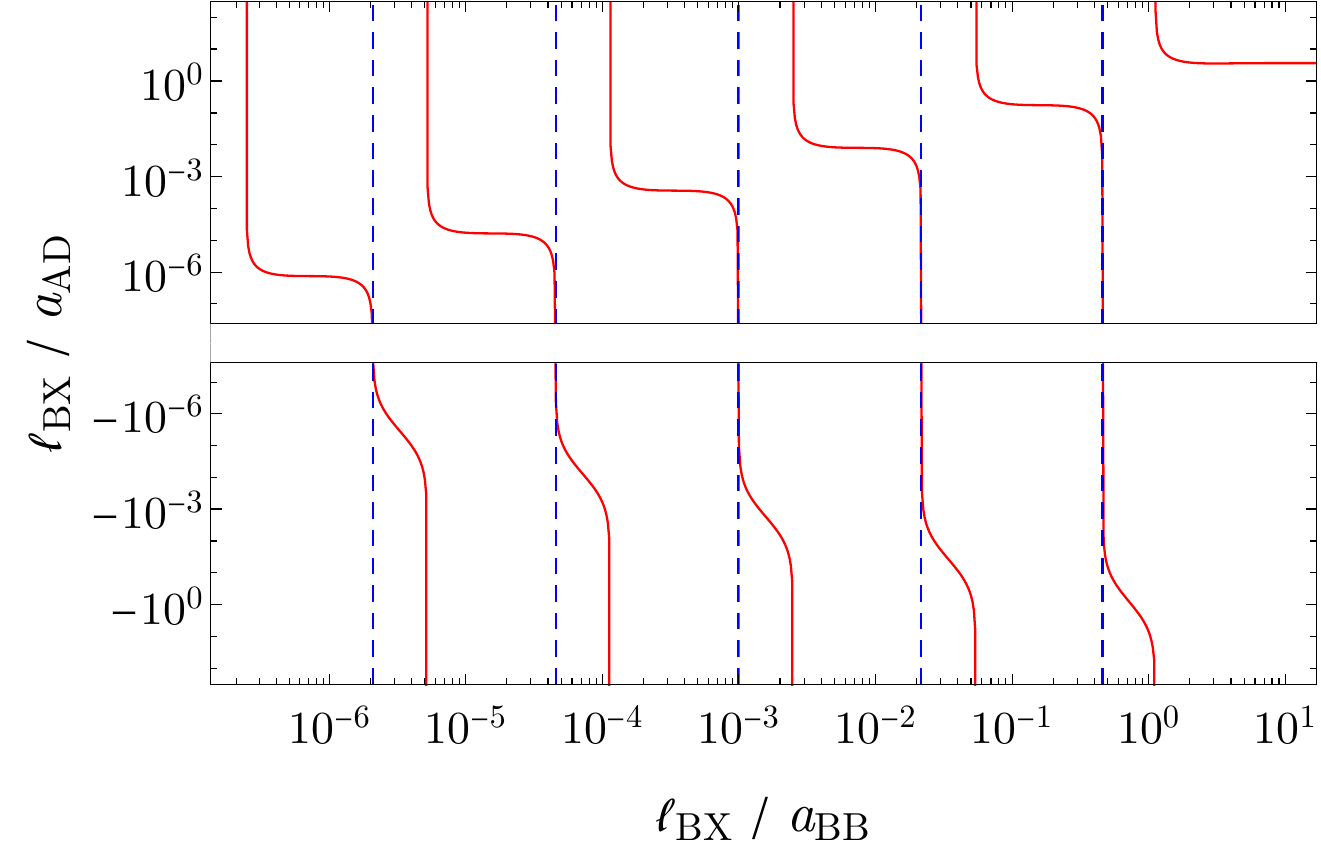}
    \caption{The atom-dimer (K-\ch{Na2}) scattering length $a_\text{AD}$ at $\abs{a_\text{BX}}\rightarrow\infty$, calculated for the lower ZR potential $U_1(R)$ with $a_\text{BB}>0$ (Fig.~\ref{fig2potcurves}(b)). The vertical dashed lines mark the values of $a_\text{BB}$ at which the lower spectrum trimers intersect the \ch{Na2} + \ch{K} threshold, as determined from the ZR data in Fig.~\ref{fig1spectrum}. Recall that this ZR calculation is equipped with a log-derivative that ties it with the FR model.}
    \label{fig4aAD}
\end{figure}
Note that the lower trimers do not extend above the \ch{Na2} + \ch{K} threshold in Fig.~\ref{fig1spectrum}, a behavior backed by analyzing the atom-dimer scattering length $a_\text{AD}$ and its dependence on $a_\text{BB}$. Figure~\ref{fig4aAD} reveals that $a_\text{AD}$ (solid) changes periodically from $\infty$ to $-\infty$ at specific $a_\text{BB}$ values that align closely with those (dashed) at which the lower spectrum loses a bound state as extracted from Fig.~\ref{fig1spectrum} (i.e., at the trimer-dimer intersections), signifying the possibility of the Efimov effect in a mixed gas of \ch{Na2} dimers and K atoms.

Similar to the homonuclear shallow dimer, the energies of the upper resonances scale inversely with $a_\text{BB}^2$ in the universal limit. Explicitly, the energies of the upper Efimov series are given by $E_n = (\ell_\text{BX}/a_\text{BB})^2 E_0 e^{-2\pi n/s_0}$, where $E_0$ is the ground-state energy at $a_\text{BB}=1\,\ell_\text{BX}$. The significantly weak coupling between the upper and lower adiabatic potentials (Fig.~\ref{fig2potcurves}(b)), arising from the large disparity between the constants $s_0$ and $s_0^*$, leads to exceptionally narrow resonance widths for the upper-spectrum states. For example, a ZR two-channel calculation shows that, at the physical Na-Na scattering length $a_\text{BB} \approx 52\,a_0$ \cite{a_Na_1,a_Na_2,a_Na_3}, the resonance width (FWHM) is only about 10\% of the corresponding state energy. 
To obtain the energies of these resonances from the FR model (i.e., as solutions of Eqs.~(\ref{IntEqs})), originally formulated for true bound states, the maximum momentum used to tabulate the matrix of Eq.~(\ref{DisIntEq}) is set to the BB dimer energy, $p_\text{max}=q_\text{max} \sim \kappa_\text{BB}$, in order to exclude the continuum states of the lower \ch{Na2} + \ch{K} channel. This momentum-space projection onto the upper channel subspace is equivalent to the real-space projection, where bound states are sought exclusively in the upper potential $U_2(R)$, ignoring nonadiabatic couplings to other channels.

\subsection{Comparison With Experiment} 
\begin{figure}[t!]
    \centering
    \includegraphics[width=\linewidth]{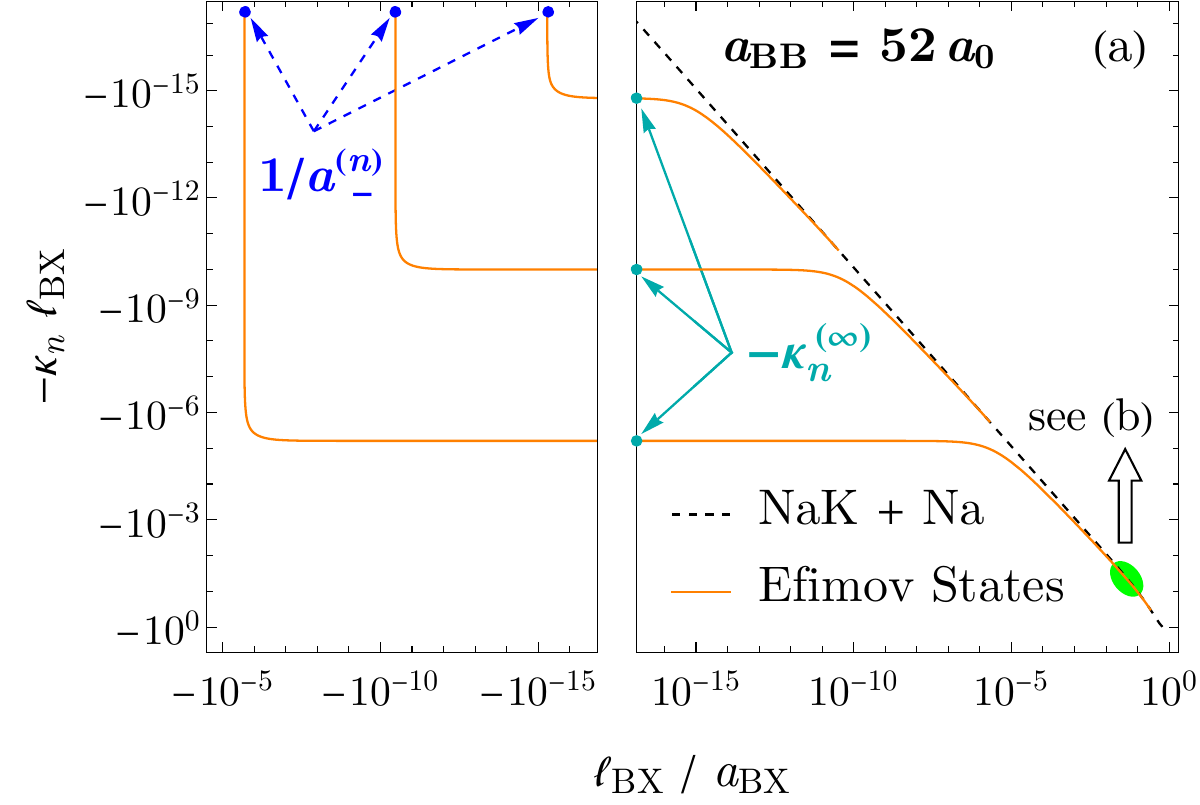}\\
    \vspace{4pt}
    \includegraphics[height=0.6\linewidth]{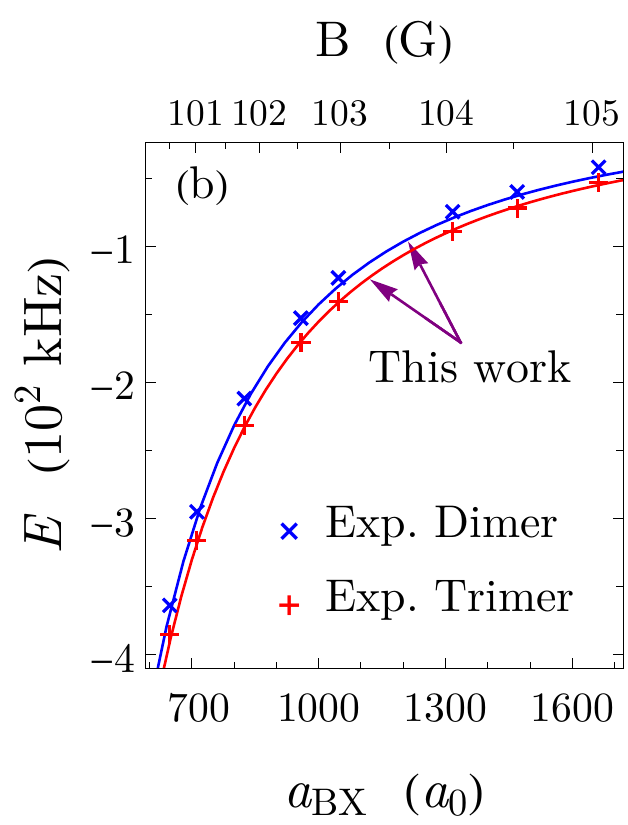}
    \hspace{2pt}
    \includegraphics[height=0.6\linewidth]{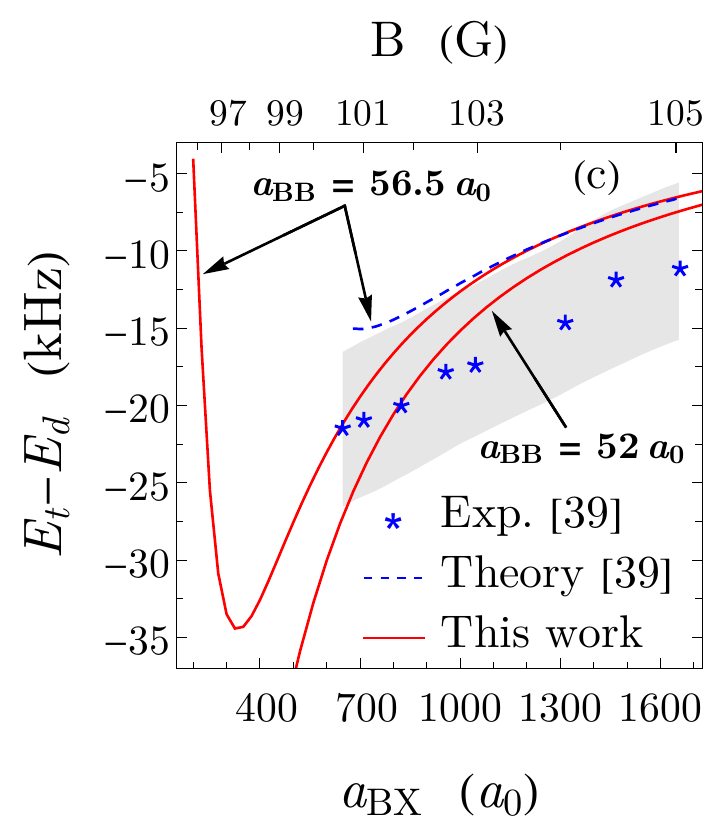}
    \caption{(a) Upper Efimov spectrum at $a_\text{BB}=52\,a_0$, computed with the FR model as a function of $1/a_\text{BX}$.
    (b) Zoomed-in view of the green-shaded region in (a), showing the binding energies of the \ch{Na2K} trimer and NaK dimer from this work (solid) compared with experimental data \cite{Zwierlein}. (c) Difference in binding energy between the trimer and dimer as predicted by the FR model (solid), and as reported theoretically (dashed) and experimentally (blue asterisks) in Ref.~\cite{Zwierlein}. The two red curves correspond to slightly different Na-Na scattering lengths used, $52\,a_0$ and $56.5\,a_0$, the latter of which is adopted in Ref.~\cite{Zwierlein}. The gray-shaded region represents the scale of uncertainty associated with experimental broadening, primarily due to magnetic field noise.}
    \label{fig5aBXspectrum}
\end{figure}

To date, the Efimov spectrum has been examined only at $\abs{a_\text{BX}}\rightarrow\infty$. Departing from unitarity in $a_\text{BX}$, the system exhibits typical Efimov behavior observed in homonuclear systems: Efimov states fade into the three-body continuum for $a_\text{BX}<0$ and merge with the dimer threshold for $a_\text{BX}>0$. For instance, in Fig.~\ref{fig5aBXspectrum}(a), the energies of the three resonances from Fig.~\ref{fig1spectrum} (upper right)  are traced in $a_\text{BX}$, with $a_\text{BB}$ set to $52\,a_0$ \cite{a_Na_1,a_Na_2,a_Na_3}. As $a_\text{BX}$ changes from $-\infty$ to $0$ with $a_\text{BB}$ fixed, Efimov states become less bound and eventually dissociate as the two-body interactions weaken in two pairs, leaving at most one pair resonant. This contrasts with the regime of Fig.~\ref{fig1spectrum}, where Efimov states never become unbound for $a_\text{BB}<0$, since two pairs remain resonant while only one pair's interaction weakens. Restated, at least two resonant pairs are required to sustain an Efimov trimer in heteronuclear systems.  

The shaded area in Fig.~\ref{fig5aBXspectrum}(a) marks the location of a recently detected \ch{^{23}Na2^{40}K} halo trimer near a broad Na-K Feshbach resonance at $\text{B}\approx110\,\text{G}$ \cite{Zwierlein}, providing a broader perspective on how the observed trimer fits within the usual Efimov scenario. Owing to the large geometric spacing in the upper spectrum, the experimental regime for such a halo trimer ($10\,\ell_\text{BX} \lesssim a_\text{BX} \lesssim 30\,\ell_\text{BX}$) lies far outside the Efimov domain in the present system ($a_\text{BX} \gtrsim 10^6\,\ell_\text{BX}$). To enable comparison with the experimental data, the theoretical dimer energy from Ref.~\cite{Zwierlein}, given as a function of magnetic field B, was matched to the FR dimer energy computed as a function of the scattering length $a_\text{BX}$, yielding the required $a_\text{BX}(\text{B})$ mapping. Figure~\ref{fig5aBXspectrum}(b) shows that the FR model prediction for the trimer energy is in strong agreement with the experimentally measured values over the examined magnetic field range. This agreement is further supported in Fig.~\ref{fig5aBXspectrum}(c), which demonstrates good consistency with both the theoretical and experimental results presented in Ref.~\cite{Zwierlein} for the trimer-dimer binding energy difference. Although a noticeable discrepancy of about 5 kHz remains between theory and experiment, it can be attributed (as explained in Ref.~\cite{Zwierlein}) to many-body effects not captured by few-body models, such as atom-dimer and atom-trimer scattering. In addition, Fig.~\ref{fig5aBXspectrum}(c) highlights the strong sensitivity to $a_\text{BB}$, as illustrated by the two red curves obtained with closely spaced values for the Na-Na scattering length. 

\subsection{The Efimov Resonances}
The unitarity (rescaled) energies $-\kappa_n^{(\infty)}$, studied in Fig.~\ref{fig1spectrum}, and the values $a_{-}^{(n)}$, linked to three-body loss resonances, are indicated for the upper spectrum by solid and dashed arrows, respectively, in Fig.~\ref{fig5aBXspectrum}(a). The current system exhibits two disparate Efimov spectra, with two sets of $a_{-}^{(n)}$ relevant for 3BR. The first resonance $a_{-}^{(0)}$ of each set serves as a 3BP, giving rise to two distinct 3BPs corresponding to the lower or upper spectra. This feature is absent in Efimov-favored systems, as increasing the mass ratio causes the two spectra to smoothly merge into a single spectrum. Specifically, as $m_\text{B}/m_\text{X}\rightarrow \infty$, $s_0-s_0^* \rightarrow 0$, and the two eigenvalues in the inset of Fig.~\ref{fig2potcurves}(b) move closer to each other, becoming more strongly coupled.

\begin{figure}[b!]
    \centering
\includegraphics[width=\linewidth]{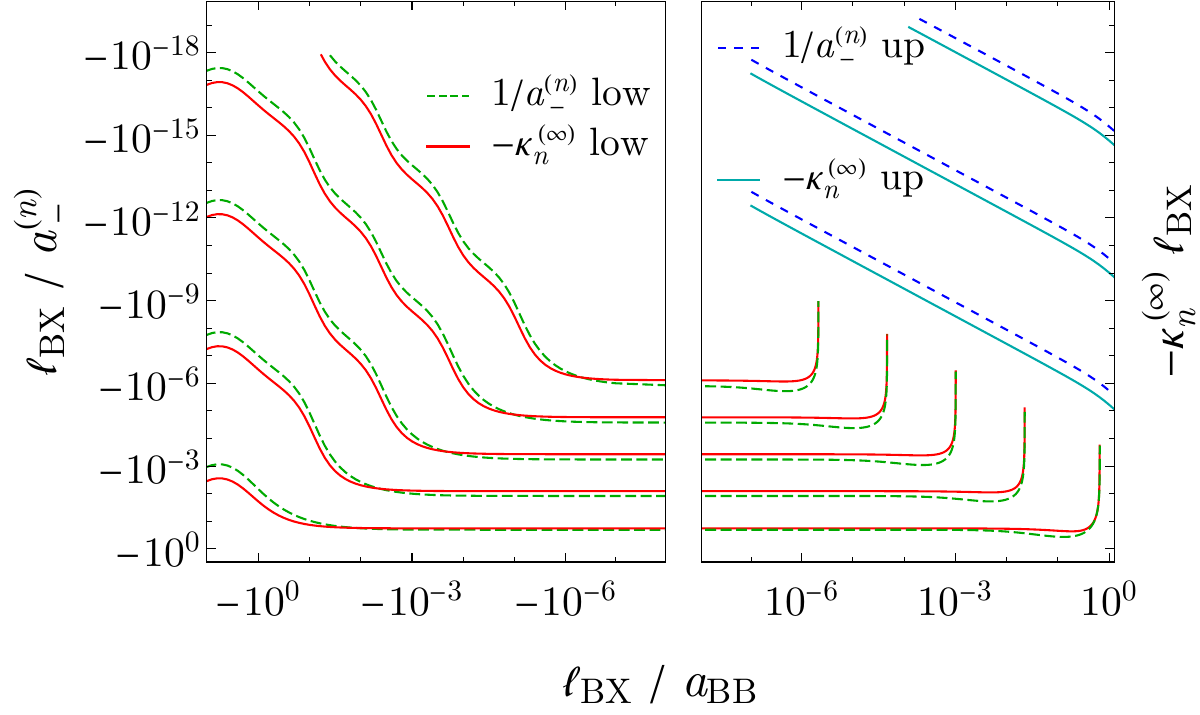}
    \caption{The Efimov resonances $a_{-}^{(n)}$ and the unitarity energies $-\kappa_n^{(\infty)}$, computed from the FR model, for the lower (green and red) and upper (blue and cyan) spectra. The presented quantities are relative to the three-body continuum, except for the lower spectrum when $a_\text{BB}>0$ (lower right), where $-\kappa_n^{(\infty)}$ (red) represent binding energies relative to the \ch{Na2} dimer energy, and $a_{-}^{(n)}$ (green) indicate the values of $a_\text{BX}$ at which each Efimov state disappears into the \ch{Na2} + \ch{K} continuum. The lower right curves were plot-truncated to avoid intersection with the upper curves.}
    \label{fig6_3BP}
\end{figure}
Given the implied connection between $a_{-}^{(n)}$ and $-\kappa_n^{(\infty)}$ through the Efimov states \cite{BRAATENReview}, $a_{-}^{(n)}$ are expected to strongly depend on $a_\text{BB}$, similarly to the unitarity energies, as illustrated in Fig.~\ref{fig1spectrum}. Using the FR model, the values $a_{-}^{(n)}$ were directly calculated over the universal range of $a_\text{BB}$ by setting the total energy to $E=0$. The dependence of $a_{-}^{(n)}$ on $a_\text{BB}$ is visualized in Fig.~\ref{fig6_3BP} for both the lower and upper spectra, with the unitarity energies $-\kappa_n^{(\infty)}$ overlaid to highlight the correlation. Manifestly, in Fig.~\ref{fig6_3BP}, $1/a_{-}^{(n)}$ are generally correlated with $-\kappa_n^{(\infty)}$. Specifically, for $a_\text{BB}<0$, $1/a_{-}^{(n)}$ display the same transition between the scaling parameters $s_0$ and $s_0^*$ as observed for the unitarity energies, leading to a significant variation in the Efimov resonance positions $a_{-}^{(n)}$ with $\abs{a_\text{BB}}$.

With $a_\text{BB}>0$, as $a_\text{BX}$ varies from $-\infty$ to $0$, the lower-spectrum states are blocked from reaching the three-body continuum by the \ch{Na2} + \ch{K} threshold (i.e., they disappear into the atom-dimer continuum). It should be emphasized that for $a_\text{BB}>0$, the lower $a_{-}^{(n)}$ in Fig.~\ref{fig6_3BP} were computed with $E$ set to the \ch{Na2} dimer energy instead of $E=0$. Thus, one should keep in mind that those lower $a_{-}^{(n)}$ would not show three-atom recombination resonances in a purely atomic gas, although there would be the possibility of recombination resonances in a mixed gas of dimers and atoms. As a result, for $a_\text{BB}>0$, the first Efimov resonance $a_{-}^{(0)}$ becomes linked to the upper-spectrum ground state, which lies five orders of magnitude (in $-\kappa^{(\infty)}$) higher than the lower spectrum (see Fig.~\ref{fig1spectrum}), leading to an exceptionally large $|a_{-}^{(0)}|$. For instance, Fig.~\ref{fig5aBXspectrum}(a) indicates that in \ch{^{23}Na2^{40}K} ($a_\text{BB}\approx52\,a_0$), the first resonance occurs at $a_{-}^{(0)}\approx26\times10^6\, a_0$, which is consistent with the value reported by Wang \textit{et al.} \cite{GreeneHetero,GreeneHeteroErratum} for \ch{^{41}K2^{87}Rb} ($a_\text{BB}\approx62\,a_0$). This effect differs from that in the \ch{Cs2Li} Refs. \cite{Ulmanis2016CsLi2,Hafner2017CsLi}, where only one loss resonance is suppressed for $a_\text{BB}>0$, and $a_{-}^{(0)}$ is modified by approximately one Efimov period, corresponding to a scaling factor $e^{\pi/s_0}\!\sim\!5$. Therefore, in light-light-heavy systems, the first resonance $a_{-}^{(0)}$ is expected to be considerably more accessible when $a_\text{BB}<0$ than when $a_\text{BB}>0$. One such example is \ch{^{85}Rb2^{133}Cs}, where the value of $a_\text{BB}\approx-390\,a_0$ leads to $a_{-}^{(0)}\approx-1800\,a_0$, attainable near the interspecies Feshbach resonances, for instance at B $\approx 107$ or $642$ G \cite{BroadFeshbachR}.

\section{Conclusion} 
In summary, we have identified two key mechanisms underlying the substantial influence of the intraspecies scattering length $a_\text{BB}$ on the Efimov scenario in Efimov-unfavored systems, as reflected in the energy spectrum and the respective resonance positions $a_{-}^{(n)}$. For $a_\text{BB}<0$, the system undergoes a gradual transition between two markedly distinct Efimov scaling parameters, resulting in a pronounced dependence of the Efimov spectrum on $\abs{a_\text{BB}}$. However, for $a_\text{BB}>0$, two qualitatively distinct Efimov spectra emerge above and below the homonuclear dimer threshold. In Efimov-unfavored mixtures (particularly LLH), these spectra are extremely separated, with the upper spectrum becoming increasingly weakly bound (i.e., shallower) as the mass ratio $m_\text{B}/m_\text{X}$ decreases. Consequently, the first Efimov resonance $a_{-}^{(0)}$ shifts to an unobservable value due to the absence of the loss resonances associated with the lower spectrum when $a_\text{BB}>0$. Crucially, both mechanisms are driven by the large disparity between $s_0$ and $s_0^*$, a distinctive feature of Efimov-unfavored systems. In addition, good correspondence with the experiment in Ref.~\cite{Zwierlein} is established by computing the binding energy of the observed \ch{Na2K} trimer while also demonstrating its relation to the Efimov regime.


\acknowledgments
This work was supported in part by NSF grants PHY-2207977 and PHY-2512984.

\appendix
\section{The Jacobi Momenta} \label{appendixA}
For three particles, the Jacobi coordinates $(\vec{p}_i,\vec{q}_i)$ describe the system's internal motion by partitioning it into a ``spectator" particle $i$ and the remaining pair $(j,k)$. The momentum $\vec{q}_i$ represents the relative motion within the $(j,k)$ pair, while $\vec{p}_i$ corresponds to the motion of the spectator particle $i$ relative to the center of mass of particles $(j,k)$. Clearly, there are three possible choices for the spectator particle $i$, and hence three corresponding coordinate sets $(\vec{p}_i,\vec{q}_i)$, which are expressed in terms of the individual particle momenta $\vec{k}_i$ as
\begin{equation} \label{Jacobi_def}
\begin{aligned}
\vec{q}_i &= \mu_i \bigg( \frac{\vec{k}_j}{m_j}-\frac{\vec{k}_k}{m_k} \bigg) \quad \text{and} \quad \\
\vec{p}_i &= \mu^i \bigg(\frac{\vec{k}_i}{m_i}-\frac{ \vec{k}_j+\vec{k}_k}{m_j+m_k} \bigg),
\end{aligned}
\end{equation}
where $i,j,k$ form a cyclic permutation of $1,2,3$. The reduced masses associated with the momenta $\vec{q}_i$ and $\vec{p}_i$ are given by
\begin{equation}
\mu_i = \frac{m_j m_k}{m_j+m_k} \quad \text{and} \quad
\mu^i = \frac{m_i(m_j+m_k)}{m_i+m_j+m_k}.
\end{equation}

This scheme is often referred to as the ``odd-man-out" notation, in which quantities are labeled by the left-out, or spectator, particle. For example, $V_1$ denotes the interaction between particles 2 and 3, $a_3$ denotes the scattering length between particles 1 and 2, etc.

\begin{figure*}[t]
    \centering
\includegraphics[height=0.221\textwidth]{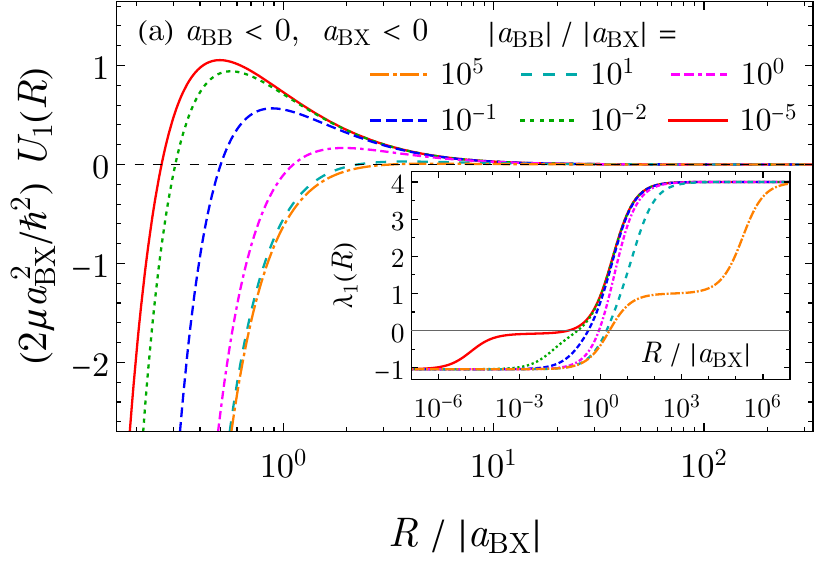}
\hspace{3pt}
\includegraphics[height=0.221\textwidth]{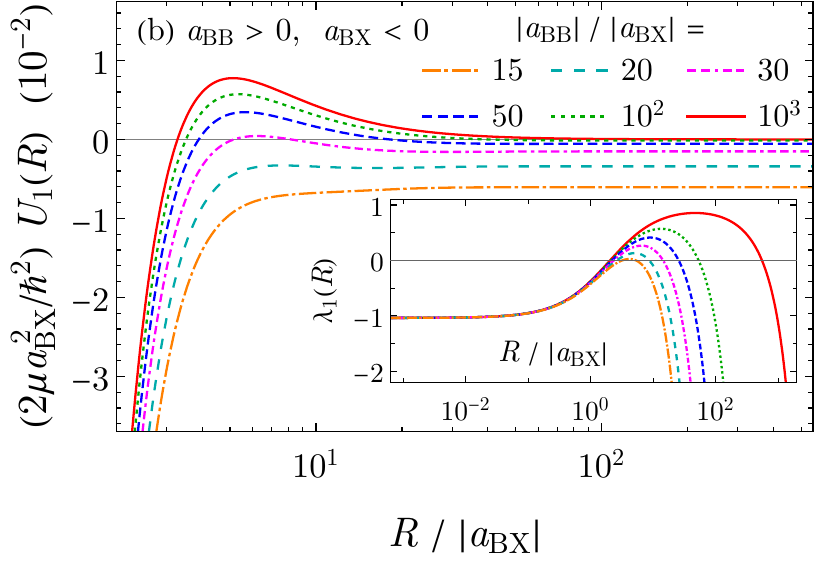}
\hspace{1pt}
\includegraphics[height=0.221\textwidth]{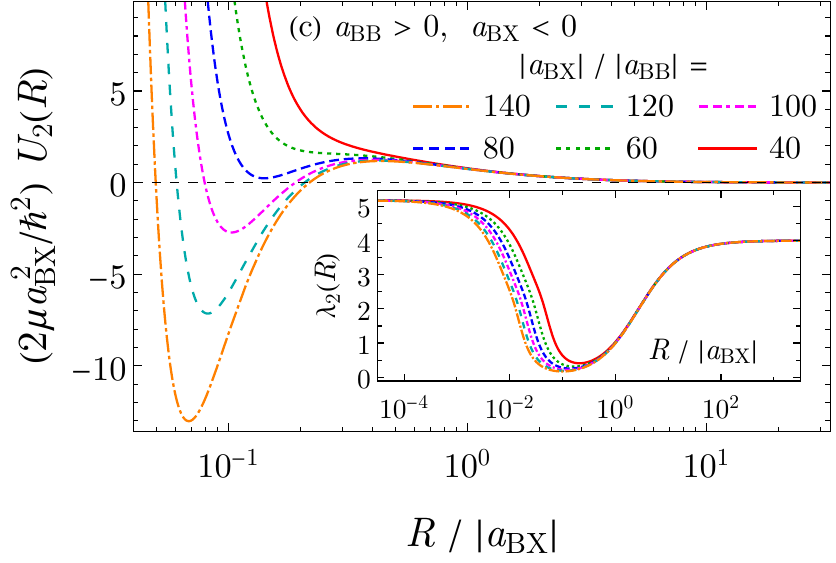}
    \caption{The zero-range hyperspherical potentials $U_n(R)$ for $a_\text{BX}<0$, shown for the lower (a and b) and upper (c) spectra. Curves within each panel correspond to different values of $|a_\text{BB}|/|a_\text{BX}|$. The insets show the corresponding hyperangular eigenvalues $\lambda_n(R)$.}
    \label{fig7_aBXpotcurves}
\end{figure*}
\section{The Zero-range Model} \label{appendixB}
The hyperangular eigenvalue $\lambda(R)$, which defines the zero-range hyperspherical potentials in Eq.~(\ref{ZRpot}), is determined by solving 
\begin{equation}
    \sum_{j=1}^{3} Z_{ij}C_j=0 \quad \text{i.e.,} \quad \det[\mathbf{Z}^{L,R}_{m_i,a_i}(\lambda)] = 0.
\end{equation}
For three distinguishable particles with total angular momentum $L$, the elements of the $3 \times 3$ matrix $\mathbf{Z}^{L,R}_{m_i,a_i}(\lambda)$ are defined as \cite{GreeneZR} 
\begin{align}
    Z_{ii}(\lambda) &= \frac{2}{\Gamma\!\left(\frac{L-\nu-1}  {2}\right)\Gamma\!\left(\frac{L+\nu+3}{2}\right)} \nonumber \\[3pt] &\hspace{1.2cm}-\frac{\sqrt{\mu/\mu_i}}{\Gamma\!\left(\frac{L-\nu}{2}\right)\Gamma\!\left(\frac{L+\nu+4}{2}\right)}\left(\frac{R}{a_i}\right), \ \text{and} \\[6pt] 
    Z_{ij}(\lambda) &= \frac{-(-1)^L}{\sqrt{\pi}\,\Gamma\!\left(\frac{2L+3}{2}\right)}\,f_{L\nu}\!\left(\arctan\left[\frac{M \mu}{m_i m_j}\right]\right),
\end{align}
for $i \neq j$, where $\nu = \sqrt{\lambda}-2$, $M$ is the total mass, and 
\begin{equation}
    \mu =\sqrt{\mu_i \mu^i}= \sqrt{\frac{m_1 m_2 m_3}{m_1+m_2+m_3}},
\end{equation} is the three-body reduced mass. The function $f_{L\nu}$ is written as
\begin{multline}
    f_{L\nu}(x)=\cos^L{x} \\ \times {}_2F_1\!\bigg(\frac{L-\nu}{2},\frac{L+\nu+4}{2};\frac{2L+3}{2};\frac{1+\cos{2x}}{2}\bigg),
\end{multline} where ${}_2F_1$ denotes a hypergeometric function. 

For a BBX system (2 bosons $+$ 1 distinguishable), the required exchange-symmetry constraint is $C_1=C_2$, which corresponds to summing the first two columns of $\mathbf{Z}^{L,R}_{m_i,a_i}(\lambda)$ and omitting the redundant second row. Hence, one obtains $\lambda$ by solving
\begin{equation} \label{BBXeq}
    \det[\begin{pmatrix}
        Z_{11}+Z_{12} & Z_{13} \\
        2Z_{31}  & Z_{33}
    \end{pmatrix}] = 0,
\end{equation} where $Z_{31}=Z_{32}$ since $m_1=m_2$. 
Now, we derive the transcendental equations satisfied by the scaling parameters $s_0$ and $s_0^*$ for spherically symmetric Efimov states ($L=0$). When all three pairs interact resonantly ($a_1=a_2=a_3 \to \infty$), Eq.~(\ref{BBXeq}) reduces to
\begin{multline} \label{s0*Eq}
    \cos(\frac{\pi \nu}{2}) \left[ \cos(\frac{\pi\nu}{2}) + {}_2F_1\!\bigg(\frac{-\nu}{2},\frac{\nu+4}{2};\frac{3}{2};\frac{u^2}{(1+u)^2}\bigg) \right]\\= 2 \times{}_2F_1\!\bigg(\frac{-\nu}{2},\frac{\nu+4}{2};\frac{3}{2};\frac{1}{2u+2}\bigg)^2.
\end{multline} 
Alternatively, when only the two heteronuclear pairs interact resonantly $(a_1=a_2\equiv a_\text{BX}\to \infty)$, while the homonuclear pair is noninteracting ($a_3 \equiv a_\text{BB} = 0$),  Eq.~(\ref{BBXeq}) becomes
\begin{multline} \label{s0Eq}
    \cos(\frac{\pi\nu}{2}) + {}_2F_1\!\bigg(\frac{-\nu}{2},\frac{\nu+4}{2};\frac{3}{2};\frac{u^2}{(1+u)^2}\bigg) = 0.
\end{multline} 
The constants $s_0^*$ and $s_0$ are then calculated by solving Eqs.~(\ref{s0*Eq}) and (\ref{s0Eq}), respectively, for a negative root $\lambda \equiv(\nu+2)^2 \to -s_0^{*2}$ (or $-s_0^2$) with an arbitrary mass ratio $u=m_\text{B}/m_\text{X}$.

\section{The Zero-range Adiabatic Potentials for $a_\text{BX}<0$}
Valuable insight into the behavior of $a_{-}^{(n)}$ (e.g., their connection with $\kappa_n^{(\infty)}$) can be gained by investigating the relevant ZR potentials. In contrast to the unitarity potentials (Fig.~\ref{fig2potcurves}), the potentials for $a_\text{BX}<0$ feature a shape barrier located roughly at $R_b \sim \abs{a_\text{BX}}$ (order of magnitude estimate), as shown in Fig.~\ref{fig7_aBXpotcurves}. As this barrier moves inward (i.e., as $\abs{a_\text{BX}}$ decreases), it overlaps with an Efimov state (initially residing in a unitarity potential), elevating the state to the relevant continuum threshold, upon which it dissociates at $a_\text{BX}=a_{-}^{(n)}$. This overlap occurs when the barrier position becomes comparable to the state's turning point at unitarity ($R_b \sim R_t$), leading to $|1/a_{-}^{(n)}| \sim \kappa_n^{(\infty)}$, thus illustrating the correlation with the unitarity energies.

However, $1/a_{-}^{(n)}$ do not precisely track the unitarity energies, i.e., the ratio $|1/a_{-}^{(n)}|/\kappa_n^{(\infty)}$ is not constant for different values of $a_\text{BB}$. For instance, $|1/a_{-}^{(n)}| > \kappa_n^{(\infty)}$ near $\abs{a_\text{BB}}\rightarrow\infty$, while $|1/a_{-}^{(n)}| < \kappa_n^{(\infty)}$ as $a_\text{BB}\rightarrow0^-$ (see  Fig.~\ref{fig6_3BP}).
This variable ratio $|1/a_{-}^{(n)}|/\kappa_n^{(\infty)}$ for $a_\text{BB}<0$ can be explained by scrutinizing Fig.~\ref{fig7_aBXpotcurves}(a), which reveals that the barrier's position and shape (width and height) hinge on the ratio $\abs{a_\text{BB}}/\abs{a_\text{BX}}$. The barrier occurs at $R_b>\abs{a_\text{BX}}$ if $\abs{a_\text{BB}} > \abs{a_\text{BX}}$ (e.g., orange and cyan), while it occurs at $R_b<\abs{a_\text{BX}}$ if $\abs{a_\text{BB}} < \abs{a_\text{BX}}$ (e.g., red and green). Hence, the condition $R_b \sim R_t$ results in $|1/a_{-}^{(n)}| > \kappa_n^{(\infty)}$ below the curve $y=-\abs{x}$ (i.e., $\abs{a_\text{BX}}=\abs{a_\text{BB}}$), and $|1/a_{-}^{(n)}| < \kappa_n^{(\infty)}$ above it. This reasoning is supported by observing that the red and green curves in Fig.~\ref{fig6_3BP} intersect (i.e., $|1/a_{-}^{(n)}| = \kappa_n^{(\infty)}$) near $\abs{a_\text{BB}}=\abs{a_\text{BX}}$, marking a transition point for the ratio $|1/a_{-}^{(n)}|/\kappa_n^{(\infty)}$. For $a_\text{BB}>0$, the barrier lies on opposite sides of $\abs{a_\text{BX}}$ for the lower and upper spectra, as evident in Figs.~\ref{fig7_aBXpotcurves}(b--c). Consequently, the two spectra lie on opposite sides of unity in this ratio, i.e., $|1/a_{-}^{(n)}|/\kappa_n^{(\infty)}>1$ for the lower spectrum and vice versa.
\bibliographystyle{apsrev4-2}
\bibliography{Ref}
\end{document}